# Inter-atom Interference Mitigation for Sparse Signal Reconstruction Using Semi-blindly Weighted Minimum Variance Distortionless Response

Ruiming Yang, Qun Wan, Yipeng Liu and Wanlin Yang
Department of Electronic Engineering
University of Electronic Science and Technology of China
Chengdu, China
{ shan99, wanqun, liuyipeng, wlyang}@uestc.edu.cn

*Abstract*—The feasibility of sparse signal reconstruction depends heavily on the inter-atom interference of redundant dictionary. In this paper, a semi-blindly weighted minimum variance distortionless response (SBWMVDR) is proposed to mitigate the inter-atom interference. Examples of direction of arrival estimation are presented to show that the orthogonal match pursuit (OMP) based on SBWMVDR performs better than the ordinary OMP algorithm.

## I. Introduction

In the last decade, the topic of sparse signal reconstruction has received a lot of attention. The goal is to find the sparsest representation of a signal by using redundant dictionary. It is NP-Hard and several suboptimal methods have been proposed [1]. Among these methods, Orthogonal Matching Pursuit (OMP) is an attractive algorithm since it is fast and easy to implement [2]. It has been proved that a signal can be reconstructed by OMP algorithm if the dictionary is incoherent and the signal is sparse enough [3] [4]. Unfortunately, experiences have shown that many dictionaries for sparse reconstruction may be highly coherent [5]. In these cases, OMP may fail to reconstruct sparse signal due to the strong inter-atom interference in highly coherent dictionary. Recently, a generalized version of OMP was proposed to mitigate this effect by introducing a sensing dictionary [6]. However, it is only applicable to noiseless measurements. In this letter, we propose a novel semi-blindly weighted minimum variance distortionless response (SBWMVDR) to mitigate the inter-atom interference in highly coherent dictionary. Different from the non-adaptive sensing dictionary design methods [6], the proposed SBWMVDR method can provide a closed-form and adaptive sensing dictionary with reduced inter-atom interference.

## II. Problem Formulation

A dictionary for the signal space $\mathbb{C}^M$ is a set of $N$ normalized vectors $\mathbf{A}_n$ ( $n \in \Omega = \{1, 2, \cdots, N\}$ ) that spans the whole space, denoted by $\mathbf{A} = [\mathbf{A}_1, \mathbf{A}_2, \cdots, \mathbf{A}_N]$. In general, the dictionary is redundant with $M < N$ and the vector $\mathbf{A}_n$ is called atom. Express a contaminated signal $\mathbf{x} \in \mathbb{C}^M$ as

$$\mathbf{x} = \mathbf{A}\mathbf{s} + \mathbf{v}, \qquad (1)$$

where $\mathbf{V}$ is a unknown additive Gaussian white noise vector with zero mean. The signal is called *K*-sparsity if *s* only contains $K$ ( $K \ll N$ ) nonzero entries. The ordinary OMP algorithm iteratively selects an atom in dictionary $A$ that correlates most strongly with the residual. At each step *k*, the best atom $\mathbf{A}_{n_k}$ is chosen as

$$n_k = \arg\max_{n \in \Omega} \hat{s}_n^{(k)}, \qquad (2)$$

where

$$[\hat{s}_1^{(k)}, \hat{s}_2^{(k)}, \cdots, \hat{s}_N^{(k)}]^T = \hat{\mathbf{s}}^{(k)} = \left|\mathbf{A}^H \mathbf{r}_k\right|, \qquad (3)$$

for $k = 1, 2, \cdots, K$, *T* denotes transpose operation and *H* presents complex conjugate transpose. We have $\mathbf{r}_1 = \mathbf{x}$ for initialization, and $\mathbf{r}_{k+1} = \mathbf{P}_k \mathbf{x}$ for $k = 1, 2, \cdots, K-1$, where $\mathbf{P}_k = \mathbf{I}_M - \hat{\mathbf{A}}^{(k)}((\hat{\mathbf{A}}^{(k)})^H \hat{\mathbf{A}}^{(k)})^{-1}(\hat{\mathbf{A}}^{(k)})^H$, $\hat{\mathbf{A}}^{(k)} = [\mathbf{A}_{n_1}, \mathbf{A}_{n_2}, \cdots, \mathbf{A}_{n_k}]$ and $\mathbf{I}_M$ is an identity matrix. As a result of the inter-atom interference, e.g, we have $\hat{s}^{(1)} = \left|\mathbf{A}^H \mathbf{x}\right| = \left|\mathbf{A}^H \mathbf{A}\mathbf{s}\right| \neq \left|\mathbf{s}\right|$ for $k = 1$. Here, the problem is how to mitigate this effect on the OMP algorithm.

## III. Generalized OMP with ideal Inter-atom Interference Mitigation

In order to identify the correct atoms in highly coherent dictionary, the generalized OMP algorithm designs a sensing dictionary *W* and uses $\hat{s}^{(k)} = \left|\mathbf{W}^H \mathbf{r}_k\right|$ rather than $\hat{s}^{(k)} = \left|\mathbf{A}^H \mathbf{r}_k\right|$

This work was supported in part by the National Natural Science Foundation of China under grant 60772146, the National High Technology Research and Development Program of China (863 Program) under grant 2008AA12Z306 and in part by Science Foundation of Ministry of Education of China under grant 109139.)

in (2) [6]. Obviously, the ordinary OMP is a special case of the generalized OMP with $\mathbf{W} = \mathbf{A}$.

A good sensing dictionary should have inter-atom interference as small as possible. In a straightforward way, we may calculate each column vector of $\mathbf{W} \in \mathbb{C}^{M \times N}$, i.e., the sensing vector $\mathbf{w}_n$, as the solution to the following minimum variance distortionless response (MVDR) problem:

$$\min_{\mathbf{w}_n} \mathbf{w}_n^H \mathbf{B}_s \mathbf{B}_s^H \mathbf{w}_n, \qquad (4)$$

$$s.\ t.\ \mathbf{A}_n^H \mathbf{w}_n = 1, \qquad (5)$$

where $\mathbf{B}_s$ consists of correct atoms corresponding to the $K$ nonzero entries of $\mathbf{s}$ in (1). The closed-form solution is given by

$$\mathbf{w}_n = \mathbf{Q}_n \mathbf{A}_n, \qquad (6)$$

for n = 1, 2, … , N, where

$$\mathbf{Q}_n = \frac{1}{\mathbf{A}_n^H \left( \mathbf{B}_s \mathbf{B}_s^H + \alpha \mathbf{I}_M \right)^{-1} \mathbf{A}_n} \left( \mathbf{B}_s \mathbf{B}_s^H + \alpha \mathbf{I}_M \right)^{-1}, \qquad (7)$$

and α is a positive regularization parameter.

When $\mathbf{A}_n$ is a column vector of $\mathbf{B}_s$, i.e., a correct atom, the minimum variance condition (4) will mitigate the correlation between the corresponding sensing vector $\mathbf{W}_n$ and other correct atoms, whereas the distortionless response constraint (5) will maintain the correlation between $\mathbf{W}_n$ and the correct atom $\mathbf{A}_n$. As a result, the nonzero entries of $s$ corresponding to the correct atoms are estimated with distortion as small as possible. On the other hand, when $\mathbf{A}_n$ is not a column vector of $\mathbf{B}_s$, the minimum variance condition (4) will prevent false atoms being selected through mitigating the correlation between $\mathbf{W}_n$ and all the correct atoms. However, the sensing dictionary $\mathbf{W}$ given by (6) is not available because the correct atoms are unknown in (7).

## IV. SBWMVDR FOR INTER-ATOM INTERFERENCE MITIGATION

Given signal $\mathbf{X}$, the probability which indicates the contribution of an atom to the reconstruction of $\mathbf{X}$ is different [5]. Like the ordinary OMP algorithm, we take the correlation between signal $\mathbf{X}$ and each atom in the dictionary as an approximate measure of this probability. Though the ordinary OMP algorithm only uses this measure to select best atom sequentially, we further exploit it to build the sensing dictionary as the solution to a semi-blindly weighted version of MVDR:

$$\min_{\mathbf{w}_n} \mathbf{w}_n^H \mathbf{C}_s \mathbf{C}_s^H \mathbf{w}_n. \qquad (8)$$

$$s.\ t.\ \mathbf{A}_n^H \mathbf{w}_n = 1\ , \qquad (9)$$

where $\mathbf{C}_s = \mathbf{AD}$, $\mathbf{D} = \text{diag}(\hat{\mathbf{s}})$ and $\hat{s} = \left| \mathbf{A}^H \mathbf{x} \right|$. The closed-form solution is given by

$$\mathbf{w}_n = \mathbf{U}_n \mathbf{a}(\boldsymbol{\theta}_n), \qquad (10)$$

where

$$\mathbf{U}_n = \frac{1}{\mathbf{a}^H(\boldsymbol{\theta}_n)\left( \mathbf{C}_s \mathbf{C}_s^H + \beta \mathbf{I}_M \right)^{-1} \mathbf{a}(\boldsymbol{\theta}_n)} \left( \mathbf{C}_s \mathbf{C}_s^H + \beta \mathbf{I}_M \right)^{-1}, (11)$$

for n = 1, … , N, and β is a positive regularization parameter.

The advantage of the sensing dictionary given by (10) is the adaptive function of inter-atom interference mitigation as a result of both the adaptive minimum interference optimization and the distortionless response constraint. Note that using $\mathbf{D} = \mathbf{I}_N$ yields non-adaptive result.

## V. SIMULATION RESULTS

We take an example of DOA estimation using single snapshot data to illustrate the performance of the generalized OMP algorithm based on SBWMVDR (SBWMVDR OMP). Consider two narrowband far-field signal sources impinge from $\varphi_1 = 8°$ and $\varphi_2 = 17°$ on a uniform linear array of 12 elements separated by half wavelength. The DOA dictionary is composed of normalized steering vectors with respect to direction grids uniformly spaced by $\Delta \theta = 0.2°$. Using the SBWMVDR, we can calculate the sensing dictionary $W$ with respect to $\mathbf{A} = [\mathbf{A}_1, \mathbf{A}_2, \cdots, \mathbf{A}_N]$, where $\mathbf{A}_n$ is the normalized steering vector corresponding to direction grid $\theta_n$ ( $0° \leq \theta_n \leq 30°$, $n \in \Omega = \{1, 2, \cdots, 151\}$ ).

Define an optimal $K$-term approximation as the solution to the following optimization problem [3]:

$$\min_{\substack{\mathbf{A}_i, \mathbf{A}_j, \mathbf{s}_i, \mathbf{s}_j \\ i, j \in \Omega}} \left\| \mathbf{x} - \mathbf{A}_i \mathbf{s}_i - \mathbf{A}_j \mathbf{s}_j \right\|_2. \qquad (12)$$

We compare the performance of the proposed method (SBWMVDR OMP) with that of the ordinary OMP algorithm, the generalized OMP with ideal (ideally generalized OMP) or non-adaptive (non-adaptive OMP) inter-atom interference mitigation, and the exhaustive 2-term approximation method. Simulation results are obtained over $J = 10000$ independent Monte-Carlo trials. Fig.1 and Fig.2 present mean absolute deviation (MAD) versus signal noise ratio (SNR), where MAD is defined as $(1/J) \sum_{j=1}^{J} \left| \hat{\varphi}_{ij} - \varphi_i \right|$, $\hat{\varphi}_{ij}$ is the DOA estimation, $i = 1, 2$. It is shown that the ordinary OMP algorithm fails to obtain accurate DOAs because the DOA dictionary is highly coherent (as high as 0.9993). The proposed method outperforms the non-adaptive OPM method and the performance is close to that of the exhaustive optimal 2-term approximation method. The ideally generalized OMP algorithm outperforms all the other methods because it exploits the correct atoms as a priori information. During our experiences, we find that the proposed method is more or less sensitive to regularization parameter. Here, we set $\beta \in [0.005, 0.5]$ according to the SNR.

## VI. CONCLUSION

In this paper, we proposed a novel SBWMVDR method to mitigate the inter-atom interference in highly coherent dictionary. Numerical simulations were provided to illustrate better performance of the generalized OMP algorithm by taking advantage of inter-atom interference mitigation based on SBWMVDR.

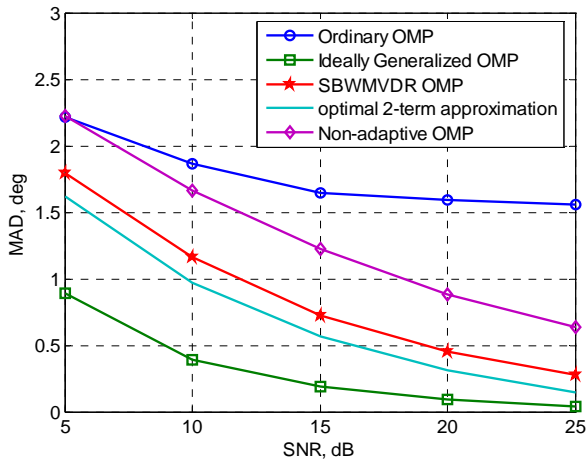

Fig.1 MAD of $\varphi_1 = 8°$ versus SNR.

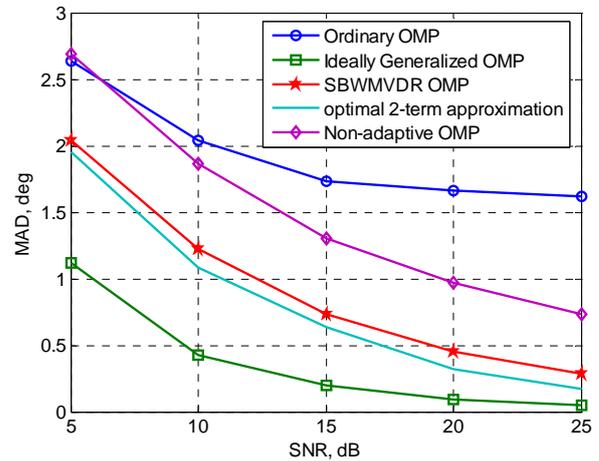

Fig.2 MAD of $\varphi_2 = 17°$ versus SNR.